\documentclass[11pt]{article}
\usepackage{epsfig}
\usepackage{amsbsy}

\begin{document}


\hyphenation{Ra-cah-expr}
\hyphenation{Ra-cah-expr-}
\hyphenation{Ra-cah-expres-sion}

\title{\b Studies of Lanthanides 6s Ionization Energy}

\author{G.\ Gaigalas${}^{}\footnote{Email: gaigalas@mserv.itpa.lt}$, Z.\ Rudzikas${}^{}$ and T.\ \v{Z}alandauskas${}^{}$, \\
        \\
        Vilnius University Research Institute of Theoretical Physics and Astronomy\\
        A.\ Go\v{s}tauto 12, Vilnius 2600, Lithuania.\\
        \\
        } 
       
\maketitle

\bigskip
\bigskip
\bigskip
\bigskip
\bigskip

\hspace{0.84cm} PACS Ref.: 03.65Ge, 31.15Ar, 31.25.Eb.

\bigskip

        
\newpage

\begin{abstract}

This work is aimed at the multi-configuration Hartree-Fock calculations 
of the $6s$ ionization energies of lanthanides with configurations $[Xe]4f^{N}6s^{2}$. 
Authors have used the ATSP MCHF version in which there are new codes for calculation 
of spin-angular parts of matrix elements of the operators of intraatomic interactions 
written on the basis of the methodology Gaigalas, Rudzikas and 
Froese Fischer, based on the second quantization
 in coupled tensorial form, the angular momentum theory in 3 spaces (orbital, spin and
 quasispin), graphical technique of spin-angular integrations and reduced coefficients (subcoefficients) 
 of fractional parentage. 
This methodology allows us to study
 the configurations with open $f$--shells without any restrictions, thus providing the possibility to 
investigate heavy atoms and ions as well as to obtain reasonably accurate values of spectroscopic data for 
such complex many-electron systems.

\end{abstract}

\newpage


\section{Introduction}

There is a considerable interest in understanding the physics and chemistry of 
heavy atoms and ions. The main problem in the investigation of such systems
is their complexity, caused by a large number of electrons and the importance of
both the correlation and relativistic effects. 
Therefore an accurate description of heavy atoms and ions requires generaly the correct treatment 
of the correlation as well as the relativistic effects. 
There are a number of approaches developed for this purpose: 
{\em configuration interaction} (CI) \cite{R97} and 
{\em multi-configuration} methods 
such as multi-configuration Hartree--Fock (MCHF) \cite{FF2}, Dirac--Fock (MCDF) methods, 
{\em many body perturbation theory} (MBPT) \cite{Lindgren:82}, etc. However the domains of their 
applicability are very different. 
Some of these methods so far may be applied only for atoms and ions having closed electronic shells or one 
electron above closed shells.

\medskip

Relativistic nature of motion implies the use of relativistic wave functions and 
relativistic Hamiltonian \cite{R97}. 
However a complete and rigorous treatment of the correlation effects 
together with the relativistic nature of the motion for heavy atoms and ions is, 
unfortunately, practically outside of today's computation possibilities.

\medskip

Fortunately, there exists a fairly large variety of atoms and their ionisation degrees, for 
which the relativistic effects are small compared to the non-spherical part of Coulomb interactions 
and, therefore, may be accurately taken into account as corrections of the order 
$\alpha^2$ ($\alpha$ is the fine structure constant) in the Pauli approximation, considered 
in details in \cite{R97}.
This is particulary true for the spectroscopic properties and processes connected with the outer 
electronic shells of an atom or ion.
Also there are some spectroscopic quantities which are described as the difference of two large numbers. 
The ionization energies belong to such category of quantities.
Relativistic effects are most important for the electrons in inner shells. 
Latters practically do not "feel" the loss of the outer electron in the process of the ionization, 
therefore the main relativistic effects cancel each other while calculating ionization energies. 
This supports the use of the approach described in this paper.
Moreover, analysis of the energy spectra of atoms considered clearly shows that the fine structure of the 
terms is really "fine", there are even no traces of splitting of a shell $f^N$ into relativistic subshells 
$f_{-}^{N_1}f_{+}^{N_2}$, typical for relativistic approach.
All this gives us the confidence that the main attention, while studying the ionization energies, must be 
paid to efficient accounting for correlation effects.

\medskip

Thus, this paper is supposed to show that some properties (such as ionization energies of valence electrons) 
of heavy atoms can be quite accurately determined using the nonrelativistic 
wave functions, accounting for correlation effects by the MCHF method and for relativistic 
effects as corrections of the order $\alpha^2$.
In addition in the paper we discribe the method of selection of the basis for the accurate  
accounting of the correlation effects important for the property under consideration,  
namely, the determination of the $6s$ ionization energies (IE) of the lanthanides.

\medskip
  
The authors were able to find only one consistent and rigorous study of 
lanthanides ionization energies \cite{SST}, including the correlation effects, performed using {\it ab initio} 
methods. In the study \cite{SST} the CI method with Gaussian-type functions was used.
This approach is typically used in molecular physics. 
The authors suppose that it is
relevant to study the ionization energies of lanthanides using the accurate methods common in
atomic physics.

\medskip

The problem in both CI and MCHF methods is to find the bases of atomic functions satisfying 
two conditions: one is to obtain accurate data and the other is to be manageable by today's 
computation possibilities. 
The right choice of the basis would allow us not only to reproduce the ionization energies
and other atomic data by {\it ab initio} methods, but it would also lead us to a better understanding 
of the importance and structure of the correlation and relativistic effects. 

\medskip

For this purpose we perform MCHF calculations 
using the multi--configuartion Hartree--Fock code from the atomic--structure package
(ATSP MCHF) \cite{FF2,F2000} in which there are new codes for
calculation of spin-angular parts of matrix elements of the operators of intraatomic 
interactions written on the basis of the methodology Gaigalas, Rudzikas 
and Froese Fischer \cite{GRFa,GRFb}, based on the second quantization in coupled tensorial form, 
the angular momentum theory in three spaces (orbital, spin and quasispin), graphical 
technique of spin-angular integrations and reduced coefficients (subcoefficients) of fractional 
parentage. The tables of such coefficients are presented in \cite{GRFb}. 
They allow us the study of configurations with 
open $f$-shells without any restrictions.
The basic concepts of our approach are presented in Section 2.

\medskip

We assume that in case of lanthanides with configurations $[Xe]4f^{N}6s^{2}$ 
the relativistic and correlation effects between the electrons of "inner shells" 
(core-core correlations) are the same for the neutral atom and ion
and then these effects (corresponding energies) cancel each other in calculation 
of {\em ionization energy} ($E_{I}$). Mean distance to the nucleus of "outer" electrons
(calculated for example by {\em single-configuration Hartree-Fock} (HF) method 
\cite{Gd,LNFK34_Er} ) is much 
larger than that of "inner" electrons. 
Therefore we expect that the correlations between "inner" and "outer" 
electrons (core-valence correlations) will be negligible. 
For the same reason we expect 
relativistic effects for "outer shells" to be not so much important as for "inner shells" 
(mean value of electron velocity is proportional to the -1 power of mean distance to the nucleus)
and they can be treated rather accurately by adding relativistic corrections to the 
non-relativistic Hamiltonian.
Then it may be possible to get quite accurate values of the ionization energies by 
MCHF approach and accounting for relativistic effects as corrections. 

\medskip

Section 3 is aimed at checking this assumption.
In Section 4 we present our final results. 
The results are compared with the previous theoretical investigations \cite{SST} and with 
values of IE compiled from experimental data \cite{Br1101,Br1666,MHRS,MZH}. 
The details of the experimental investigation 
of the ionization energies of lanthanides can be found in \cite{Camus,Smith,W}.
Section 5 serves for conclusions.



\section{Approach}
We define the ionization energy as $E_{I} = E_{ion} - E_{g}$, where 
$E_{g}$ and $E_{ion}$ are the ground state energies of neutral 
and singly ionized atoms correspondingly. The ground state of 
a neutral lanthanide atom is 
\begin{equation}
1s^{2}2s^{2}2p^{6}3s^{2}3p^{6}3d^{10}4s^{2}4p^{6}4d^{10}5s^{2}5p
^{6}4f^{N}6s^{2}~~\equiv [Xe]4f^{N}6s^{2}
\end{equation}

\medskip

and that of singly ionized one 
\begin{equation}
1s^{2}2s^{2}2p^{6}3s^{2}3p^{6}3d^{10}4s^{2}4p^{6}4d^{10}5s^{2}5p
^{6}4f^{N}6s^{1}~~\equiv [Xe]4f^{N}6s^{1}.
\end{equation}

Here $N$ corresponds to $3,...,7$ for $Pr,...,Eu$ and to $9,...,14$ for $Tb,...,Yb$.

\bigskip

In our calculations we account for the relativistic effects by the following relativistic 
shift operator (notations for ${\cal H}_i$ are taken from \cite{R97}):

\begin{equation}
\label{RelCor}
{\cal H}_{RelCor} = {\cal H}_1 + {\cal H}_2 + {\cal H}_3 + {\cal H}^{\prime}_{5} + {\cal H}_{mp}.
\end{equation}

\medskip

Here the {\em mass correction} term ${\cal H}_{1}$ and {\em orbit--orbit} 
term ${\cal H}_{2}$ are given by

\begin{equation}
\label{H12}
{\cal H}_{1}=-\frac{\alpha ^2}8\sum_{i=1}^N{\bf p}_i^4, 
\qquad \mbox{}\qquad 
{\cal H}_{2}=-\frac{\alpha ^2}2\sum_{i<j}^N\left[ \frac{\left({\bf p}_i\cdot 
{\bf p}_j\right) }{r_{ij}}+
\frac{\left({\bf r}_{ij}\cdot\left( {\bf r}_{ij}\cdot {\bf p}_i\right)
{\bf p}_j\right) }{r_{ij}^3}\right].
\end{equation}


The ${\cal H}_{3}$ stands for the one-particle (${\cal H}^{\prime}_{3}$) and 
two-particle (${\cal H}^{\prime\prime}_{3}$) {\em Darwin terms}. 
They are given by

\begin{equation}
\label{H3}
{\cal H}_3={\cal H}^{\prime}_3 + {\cal H}^{\prime\prime}_3
=\frac{Z \alpha ^2 \pi}2\sum_{i=1}^N{\boldmath{\delta }}\left({\bf r}_i\right)
 -\pi \alpha ^2\sum_{i<j}^N{\boldmath{\delta }}\left( {\bf r}_{ij}\right) ,
\end{equation}

and {\em spin--spin contact} term ${\cal H}^{\prime}_{5}$ is
 
\begin{equation}
\label{H5p}
{\cal H}^{\prime}_5=-\frac{8\pi \alpha ^2}3\sum_{i<j}^N\left( {\bf s}_i\cdot 
{\bf s}_j\right) \delta \left( {\bf r}_{ij}\right).
\end{equation}

The operators (\ref{H12}-\ref{H5p}) are of the order $\alpha^2$.

\medskip

The {\em mass--polarization} correction term ${\cal H}_{mp}$ is given by

\begin{equation}
\label{Hmp}
{\cal H}_{mp} = -\frac{1}{M} \sum_{i < j}\left({\bf p}_i\cdot {\bf p}_j\right).
\end{equation}

The expressions (\ref{H12}-\ref{Hmp}) are presented in atomic units. 

\medskip

We expect the operator (\ref{RelCor}) to enable us to take into account
the main relativistic corrections to ionization energy.

\bigskip

For the calculation of ionization energy we used MCHF method. 
In this approach, the atomic state function $\Psi (\gamma LS)$ is expressed as a linear 
combination of {\em configuration state functions} (CSFs) $\Phi (\gamma_{i} LS)$, i.e. 

\begin{equation}
\Psi (\gamma LS) = \sum_{i} c_{i} \Phi (\gamma_{i} LS).
\end{equation}

A set of orbitals, or {\em an active set}
(AS), determines the set of all possible CSFs or the {\em complete 
active space} (CAS) for MCHF calculation. The size of the latter grows rapidly 
with a number of electrons and also with the size of the orbital AS. 
Most MCHF expansions are therefore limited to a {\em restricted active 
space} (RAS) \cite{FF2}. 
The RAS is spanned by all CSFs that can be generated from a given active set, of orbitals, 
with some constrains. The constrains are derived from the notions of different types of correlations 
discussed below. No "relaxation" effects were included.

\medskip

For complex atoms and ions, a considerable
part of the effort must be devoted to integrations over spin--angular variables,
occurring in the matrix elements of the operators under consideration. 
In the papers \cite{R97,GRFa,Gaigalas:99}, an efficient approach for finding matrix 
elements of any one-- and two--particle operator between complex
configurations is suggested. 
It is based on the extensive exploitation of the symmetry properties of the quantities 
of the theory of complex atomic spectra, presented in the secondly quantized form, 
in orbital, spin and quasispin spaces. 
It is free of shortcomings of previous approaches. 
This approach allows one to generate fairly accurate databases of atomic parameters
\cite{FroeseYG:1995,FroeseG:1996} and will be used in our paper.

\medskip

According to the approach by \cite{GRFa,Gaigalas:99},
a general expression of submatrix element 
for any scalar two--particle operator between functions with $u$ open shells,
valid for both non--relativistic and relativistic wave functions, 
can be written down as follows:
\begin{eqnarray}
\label{eq:matrix-d}
\lefteqn{
   \left( \psi _u\left( LS\right) \left\|
   \widehat{G}^{\left( \kappa _1\kappa _2k,\sigma_1\sigma _2k\right) }
   \right\| \psi _u\left( L^{\prime }S^{\prime }\right) \right) }
   \nonumber  \\[1ex]
   & & = \displaystyle {\sum_{n_il_i,n_jl_j,n_i^{\prime }l_i^{\prime },n_j^{\prime
   }l_j^{\prime }}}
   \left( \psi _u\left( LS\right) \left\| \widehat{G}
   \left( n_il_i,n_jl_j,n_i^{\prime }l_i^{\prime },n_j^{\prime }
   l_j^{\prime } \right)
   \right\| \psi _u\left( L^{\prime }S^{\prime }\right) \right)
   \nonumber  \\[1ex]
   & & = \displaystyle {\sum_{n_il_i,n_jl_j,n_i^{\prime }l_i^{\prime },n_j^{\prime
   }l_j^{\prime }}}\displaystyle {\sum_{\kappa _{12},\sigma _{12},\kappa
   _{12}^{\prime },\sigma _{12}^{\prime }}}\sum_{K_l,K_s} \left( -1\right) ^\Delta \Theta
   ^{\prime }\left( n_i\lambda _i,n_j\lambda _j,n_i^{\prime }\lambda _i^{\prime
   },n_j^{\prime }\lambda _j^{\prime },\Xi \right) 
   \nonumber  \\[1ex]
   & & \times T\left(
   n_i\lambda _i,n_j\lambda _j,n_i^{\prime }\lambda _i^{\prime },n_j^{\prime
   }\lambda _j^{\prime },\Lambda ^{bra},\Lambda ^{ket},\Xi ,\Gamma \right)
   R\left( \lambda _i,\lambda _j,\lambda _i^{\prime },\lambda _j^{\prime},
   \Lambda ^{bra},\Lambda ^{ket},\Gamma \right),
\end{eqnarray}
where $\Gamma$ refers to the array of coupling parameters connecting the recoupling
matrix 

$R\left( \lambda _i,\lambda _j,\lambda
_i^{\prime },\lambda _j^{\prime },\Lambda ^{bra},\Lambda ^{ket},\Gamma
\right) $ to the submatrix element 

$T\left(
n_i\lambda _i,n_j\lambda _j,n_i^{\prime }\lambda _i^{\prime },n_j^{\prime
}\lambda _j^{\prime },\Lambda ^{bra},\Lambda ^{ket},\Xi ,\Gamma \right) $, 
$\lambda_i \equiv l_is_i$, parameter $\Xi$
implies the array of coupling parameters that connect $\Theta $
to the tensorial part, $\Lambda ^{bra}\equiv \left( L_iS_i,L_jS_j,L_i^{\prime }S_i^{\prime
},L_j^{\prime }S_j^{\prime }\right) ^{bra}$ is the array for the bra
function shells' terms, and similarly for $\Lambda ^{ket}$.
The expression (\ref{eq:matrix-d}) has summations over intermediate ranks
$\kappa _{12}$, $\sigma _{12}$, $\kappa _{12}^{\prime }$, $\sigma _{12}^{\prime }$, 
$K_l$, $K_s$ in
$T\left(
n_i\lambda _i,n_j\lambda _j,n_i^{\prime }\lambda _i^{\prime },n_j^{\prime
}\lambda _j^{\prime },\Lambda ^{bra},\Lambda ^{ket},\Xi ,\Gamma \right) $.

\medskip

In calculating the spin--angular parts of a submatrix element using
(\ref{eq:matrix-d}), one has to compute the following quantities
(for more details see \cite{GRFa}:

\begin{enumerate}
\item  The recoupling matrix $R\left( \lambda _i,\lambda _j,\lambda
_i^{\prime },\lambda _j^{\prime },\Lambda ^{bra},\Lambda ^{ket},\Gamma
\right) $. This recoupling matrix accounts for the change in
going from matrix element

$\left( \psi _u\left( LS\right) \left\|
\widehat{G} \left( n_il_i,n_jl_j,n_i^{\prime }l_i^{\prime },n_j^{\prime }
l_j^{\prime } \right) \right\| \psi _u\left( L^{\prime }S^{\prime }\right) \right)$,
which has $u$ open shells in the {\em bra} and {\em ket} functions, to the submatrix
element

$T\left(
n_i\lambda _i,n_j\lambda _j,n_i^{\prime }\lambda _i^{\prime },n_j^{\prime
}\lambda _j^{\prime },\Lambda ^{bra},\Lambda ^{ket},\Xi ,\Gamma \right) $,
which has only the shells being  acted upon by the two--particle operator in its 
{\em bra} and {\em ket}
functions.

\item
The submatrix element
$T\left( n_i\lambda _i,n_j\lambda _j,n_i^{\prime }\lambda _i^{\prime
},n_j^{\prime }\lambda _j^{\prime },\Lambda ^{bra},\Lambda ^{ket},\Xi
,\Gamma \right) $ for tensorial products of creation/annihilation operators that act
upon a particular electronic shell. So, all the advantages of tensorial algebra and
quasispin formalism may be efficiently exploited in the process of their calculation.

\item  Phase factor $\Delta $.

\item  $\Theta ^{\prime }\left( n_i\lambda _i,n_j\lambda _j,n_i^{\prime
}\lambda _i^{\prime },n_j^{\prime }\lambda _j^{\prime },\Xi \right) $, which
is proportional to the radial part and corresponds to one of
$\Theta \left(
n\lambda ,\Xi \right) $,...,$\Theta \left( n_\alpha \lambda _\alpha ,n_\beta
\lambda _\beta ,n_\gamma \lambda _\gamma ,n_\delta \lambda _\delta ,\Xi
\right) $. It consists of a submatrix element
 $\left( n_i\lambda
_in_j\lambda _j\left\| g^{\left( \kappa _1\kappa _2k,\sigma _1\sigma _2k\right)
}\right\| n_i^{\prime }\lambda _i^{\prime }n_j^{\prime }\lambda _j^{\prime }\right)
$, and in some cases of simple factors and 3$nj$--coefficients.
\end{enumerate}
The abovementioned method of the definition of spin--angular parts becomes especially important 
in the investigation of the complex systems in both relativistic and nonrelativistic approaches.

\medskip

The usage of MCHF as well as MCDF methods gives accurate results only when the RAS is formed properly.
Therefore the next chapter is dedicated to the analysis of this problem. 


\section{RAS construction}

Large scale systematic MCHF calculations (except for Er \cite{LNFK34_Er} and Gd \cite{Gd}) 
of $E_{I}$ of lanthanides have not been done yet. 
Therefore, following the methodology of \cite{FF2}, it is important to investigate 
the structure of ground configurations, to impose the {\em core} and 
{\em valence} shells and to 
evaluate {\em valence--valence} (VV), {\em core--valence} (CV) and  {\em core--core} (CC) 
correlations.

\medskip

It is always a question when we can assume that a shell is a part of the core, and 
when it should be treated as a valence shell. The answer is not trivial even for 
boron like ions, and in our case it is even more complicated because of the complexity of 
configurations under consideration. Our purpose is to take care of the correlation 
effects that do not cancel each other between ion and atom. 

\medskip

In this chapter we will discuss some practical possibilities of 
RAS construction using an example of Er \cite{LNFK34_Er}.

\subsection{Single-configuration HF calculations}

We can get the first insight into the structure of Er and Er$^{+}$ ground states from 
the single-configuration HF calculations. 
The resultant ground state energies and mean values of various operators 
of $nl$ radial functions are presented in Table I. 
Resultant energies are practically the same as those presented in \cite{FF77, TSSMK}. 

\medskip

The fact that the mean values of $<r>$, $<r^2>$ operators are much higher 
and at the same time the mean value of $<1/r>$ is much smaller 
for $6s$ function than those for $5s$, $5p$ and $4f$ functions 
shows that the $6s$ function is much more remote from the nucleus than the others. 

\medskip

Similar analysis shows that the open $4f$ shell is closer to the nucleus than 
$5s$ and $5p$. 

\medskip

The same situation remains for the Er$^{+}$ ion (corresponding values presented in brackets).   
Therefore, we have a difficulty in treatment of "outer" electrons: usually as 
outer (valence) shells the open ones are considered, but sometimes the closed shells 
($6s^2$ in our case) are included too \cite{FF2}. 
For the light atoms these shells are spatially "outer". 

\medskip

The same qualitative picture is valid for other lanthanides considered.

\medskip

It is interesting to notice that $2p$ and $3p, 3d$ electrons are spatially closer to the
nucleous than respectively $2s$ and $3s$. This fact may be explained by the increasing role of 
relativistic effects for inner electrons in heavy atoms, which may need already 
proper account for so called indirect relativistic effects. 


\subsection{Core $I$} 

In this case we use the core 

$I$. $[Xe]$ ${}^{1}S$ 

and we treat $4f$, $6s$ as valence shells. 
We treat $4f$ shell as a valence shell because it is open and $6s$ - because 
the corresponding radial function is much more remote from the nucleus than others. 
This approach is close to the advices given in \cite{FF2}. 

\medskip

The basis for the MCHF expansion was formed using the CSFs of the configurations made 
of single and double (S, D) excitations from the valence shells to some {\em destination set}.
There were two types of destination sets used:


\begin{equation}
\label{core_a}
a =  \left \{ 5d,5f,5g,6p,6d\right\},
\end{equation}


\begin{equation}
\label{core_b}
b = a + \left \{6f,6g,6h,7s,7p,7d\right\}.
\end{equation}


Further on we denote the basis as a core with subscript of destination set.
For example, $I_{a}$ denotes the basis, consisting of CSFs of the configurations, 
made by S, D excitations from  $4f^{12}6s^{2}$ for Er and $4f^{12}6s^{1}$ for $Er^{+}$ to 
the destination set "$a$" and cores $[Xe]$.  
The numbers of CSFs in the bases (NCSF) are presented in Table II.  

\medskip

The weight for the main CSF was found to be 0.977 for $I_{a}$ (and similar for $I_{b}$). 
This value is close to that (0.949) found by CI method \cite{SST}.  
The mean distances of radial functions from the nucleus are found to be up to 2 \% 
smaller than those for single-configuration HF calculations. 
For example, $<r>_{4f}$ = 0.752 a.u. for $I_{a}$  (0.748 a.u. for $I_{b}$)
and $<r>_{6s}$ = 4.550 a.u. for $I_{a}$  (4.534 a.u. for $I_{b}$).


\subsection{Cores $II$, $III$} 

\medskip

In this case, only $6s$ is treated as a valence shell, because of its spatial location. 
We expect this strategy to be more efficient for the calculations of $6s$ ionization energy
because as we can see from single-configuration HF calculations the mean distance of $4f$ 
radial functions is not much different for Er and Er$^{+}$. 
As the cores we use


$II$. $[Xe]4f^{12}$ with not fixed term 

and

$III$. $[Xe]4f^{12}$ with fixed term $^{3}H$. 

There were five types of destination sets used with these cores, namely, 

(\ref{core_a}) and (\ref{core_b}) as for core $I$ and three more

\begin{equation}
\label{core_c}
c = b + \left \{7f,7g,7h,7i,8s,8p,8d\right\},
\end{equation}

\begin{equation}
\label{core_d}
d = c + \left \{ 8f,8g,8h,8i,8k,9s,9p,9d\right\}, 
\end{equation}

\begin{equation}
\label{core_e}
e = d + \left \{9f,9g,9h,9i,9k,9l,10s,10p,10d\right\}.
\end{equation}

The results of MCHF calculations (Er and Er$^{+}$ ground state energies and 
ionization energies) are also presented in Table II.
The weights of the main CSFs in MCHF expansions are between 0.960 -- 0.980 for all bases with 
cores II, III.
The mean distance from the nucleus for $6s$ radial function is greater than the one
obtained from single-configuration HF calculations but smaller than that obtained using 
the bases with core $I$.
For example,  $<r>_{6s}$ = 4.560 a.u. for $III_{a}$, 4.564 a.u. for $III_{b,d,e}$.

\medskip

Here we would like to draw an attention to the fact that in order to accurately account for the 
correlation effects of some type (e.g. core-core or core-valence) the destination set 
should be big enough. 
In the calculation of the ionization energy it is especially 
important to properly accout for the correlation effects of the same nature for an atom and ion.
For example, the destination sets $a$ for the cores $II$ and $III$ are too small and 
therefore lead to the far from true values in the ionization energy because the number of CFSs 
in the ions MCHF expansion is too small. 
It becames particularly obvious in the case of $III_a$ for Ho for which the value of 
ionization energy $E_I=19.189$ (see Table II) is far from the real one.
But the increase of the destination set up to the $b$ already gives 
balanced inclusion of the correlation effects for an atom and ion and reasonable  
values of IE. 
The further increse of destination set gives the convergence of the IE to the value 
defined by the choice of the core, and the approach (Hamiltonian).

\subsection{Strategy of RAS formation} 
\label{RASformation}

As we can see from Table II, the basis formed with the same destination sets is 
the biggest for core $I$, the medium for core $II$ and the smallest for core $III$.
Correspondingly the energies are the lowest for core $I$, 
the medium for core $II$ and the highest for core $II$. 
This means that the bases of core $I$ account for more correlation 
effects, than the ones of cores II and III.
Nevertheless the ionization energies obtained using cores $II$ and $III$ are practically the same, and the 
ones obtained using core $I$ are much worse. 
This is due to the fact that the basis formed using the destination set "b" for core $I$ is 
not enough to account for the correlation effects of $4f$ electrons, which, represented 
in full, cancel between Er and Er$^{+}$.

\medskip

So, the most efficient strategy is to use the MCHF expansions with a frozen core of 
the type $[Xe]4f^{N}$ $^{2S+1}L$ and single, double excitations from 6s.
This strategy was used when forming the bases for $E_{I}$ calculations of other lanthanides. 
Corresponding sizes of the bases are simillar to the ones for Er. 
For example, the bases of the type similar to $III_{e}$ consisted of 3018 CSFs for Pr, Nd, Dy, Ho, of 
2938 CSFs for Pm, Tb and of 2240 CSFs for Sm, Tm.


\section{6s ionization energy}

\subsection{Nonrelativistic}
\label{Nonrelativistic}

The nonrelativistic $6s$ ionization energies of atoms considered are presented in Table III. There $E_{I}$ stands
for a value of ionization energy calculated by MCHF method, Exp - experimental results
\cite{MZH, W}. 
For comparison we also present single-configuration HF and $CI$ \cite{SST} results.
We were not able to obtain the relevant result for Europium due to the problems with 
the convergence of MCHF equations. 

\medskip

Fig. 1 shows Z dependence of ionization energies calculated by single-configuration 
HF, CI, MCHF methods and the experimental one.

\medskip

The differences between the MCHF energies of the ground states and the ones obtained by 
single-configuration HF method ($\Delta E$) for all Z vary from 0.626 eV to 0.707 eV. 
Their absolute value is smaller than that predicted in \cite{SST}. So, in general 
we encounter less correlation effects for the ground states.
For example, for Er in \cite{SST} there is $\Delta E$=-15.339 eV and our value is 
$\Delta E$=-0.669 eV.

\medskip

Nevertheless our computed values of ionization energies are 
closer to the experimental ones
than CI \cite{SST} (see the root-mean-square deviations $\sigma$ of calculated results from 
experimental measurements in Table III). 
For example, for Er in \cite{SST} there is 5.077 eV and our value is 5.792 eV whereas 
the experimental value is 6.108 eV \cite{MZH, W}.
So, though we account for less correlation effects in general, however we get better value of 
ionization energy because we account for more correlation effects that do not cancel 
between the atom and the ion. 

\medskip
 
For the smaller Z the results of CI and MCHF calculations are quite close. For example, for 
Pr (Z=59) the difference between CI and MCHF values is only 0.019 eV (i.e. less than 1\%). 
Meanwhile the MCHF results grow faster with the increasing Z and for large Z they are 
much more closer to the experimental ones. For example, for  Tm (Z=69) 
the difference between CI and MCHF values of $E_{I}$ grows up to 0.982 eV (i.e. 16\%). 

\medskip

Fig. 2 shows the Z dependence of the influence of correlation effects $\Delta E$ on the $E_{I}$  
calculated by CI method with Davidson Q correction ($CI_{+Q}$) \cite{SST} and by MCHF method. 
Davidson Q correction is supposed to aproximately account for the higher order correlation effects. 
We define the influence as $\Delta E_{I}=E_{I}-E_{IHF}$, where $E_{IHF}$ stands for the value of 
ionization energy calculated by single-configuration HF method.

\medskip

As we see in Fig. 2, the values of $\Delta E_{I}$ calculated by MCHF and by $CI_{+Q}$ methods show 
different Z-behaviour. While $CI_{+Q}$ results tend to decrease with Z the MCHF ones 
increase. 
We expect the increase of influence of correlation effects with Z to be a real one because 
of two reasons: the MCHF results are closer to the experimental ones and it is more 
realistic to expect that with increasing number of electrons the influence of 
their correlation effects grows too.

\subsection{With relativistic corrections}
\label{relativistic}

The $6s$ ionization energies calculated with various relativistic corrections are
presented in Table IV. 
There $E^{i}_{I}$ stands for a value of ionization energy calculated by MCHF method using 
nonrelativistic Hamiltonian with relativistic corrections ${\cal H}^{i}$. 
Here $i=1,2,3$ and ${\cal H}^1 ={\cal H}_1 + {\cal H}_3 + {\cal H}^{\prime}_5$, 
${\cal H}^2={\cal H}^1 + {\cal H}_{mp}$, ${\cal H}^3={\cal H}^2 + {\cal H}_{2}$. 

\medskip

For comparison we also present ionization energies calculated using the Relativistic 
Hartree-Fock method (RHF), 
the ones of CI with Davidson Q correction and estimated relativistic corrections ($CI_{Est}$) \cite{SST} 
(these values practically cannot be considered as {\em ab initio}) 
as well as the experimental results \cite{MZH, W} (Exp). 

\medskip

Two--electron relativistic corrections ${\cal H}_2$, ${\cal H}^{\prime\prime}_3$ and 
${\cal H}^{\prime}_5$ are generally of the same order of magnitude, but 
their contribution may have different signs, therefore they all must be taken into account simultaneously. 
Therefore the results $E^{3}_{I}$ in Table IV must be considered as the most correct, 
in spite of the fact that the data of the columns $E^{1}_{I}$, $E^{2}_{I}$ seem to be slightly closer 
to the experimental ones.
The point is that one--electron operators ${\cal H}_1$ and ${\cal H}^{\prime}_3$ 
have large contributions but of opposite signs 
therefore are very sensitive to the accuracy of the wave functions used.

\medskip

The results of Table IV also suggest that accounting for relativistic effects as relativistic 
corrections of the order $\alpha^2$ usually improves the ionization energies of rare earths 
(compare with $E_I$ column of Table III), 
but there may occur cases (for example Ho, Tm) where such an improvement worsens the final result.
Therefore taking into consideration the relativistic effects for heavy atoms having open $f$--shells 
requires further studies.


\medskip

The results presented in the subsections \ref{Nonrelativistic} and \ref{relativistic} 
show that our values of ionization energies are the closest to the experimental ones with respect to other ones 
obtained by pure {\em ab initio} methods and in most cases are even better than the ones 
obtained by using semiempirical corrections 
in spite of the fact that  
the RAS is formed in such a way that the coresponding bases are relatively small. 
The results obtained allow to evaluate more precisely the influence of correletion effects 
to the ionization energies of the $6s$ electrons.

\medskip

The results of \ref{relativistic} subsection show that the relativistic effects accounted in the form 
of (\ref{RelCor}) in MCHF approach are not appropriate for the elements Ho, Tm.
The values of their ionization energies with the corresponding corrections are bigger than experimental ones.

\medskip

The strategy of RAS formation presented in subsection \ref{RASformation} 
gives a hint for the formation of corresponding bases in relativistic approach too.
The bases ($III_a$ -- $III_e$) presented in the subsection \ref{RASformation} contain 
the minimum number of CSFs but the correlation effects are adequately accounted for an atom and ion.
So such bases (but with the relativistic splitting of subshells) should be used for the corresponding 
study by relativistic MCHF method as well.


\section{Conclusion}

The results obtained show that if the correlation effects of inner shells cancel each 
other between atom and ion, then it is possible to get quite accurate data 
on ionization energies by MCHF method 
accounting for the correlation effects of the outer electrons only. 
This assumption takes place in the case of ionization energy of lanthanides with 
configurations $[Xe]4f^{N}6s^{2}$.  

\medskip

Our results on 6s ionization energy of lanthanides with configurations $[Xe]4f^{N}6s^{2}$ 
are more accurate than the data found using the $CI$ method \cite{SST}.  



\medskip

The influence of the correlation efects on the ionization energy of
lanthanides with configurations $[Xe]4f^{N}6s^{2}$ is higher than it has been found before 
\cite{SST} and this influence grows with Z (with N).
However, the convergency of the value studied to true one with the increase of the basis 
is often not smooth. This statement is illustrated very well by the intermediate value 
of $E_I$ for Ho 19.189 eV (basis $III_a$ in Table II).  

\medskip

The results presented demonstrate the ability of the approach by Gaigalas {\it et al.} \cite{R97,GRFa,GRFb} 
based on the second quantization in coupled tensorial form, the graphical technique of spin--angular 
integration, quasispin formalism and reduced coefficients (subcoefficients) of fractional parentage   
to obtain reasonably accurate data on the ionization energies of heavy atoms and ions, having open $f$-shells.

\medskip

Acounting for the relativistic effects as the corrections of order $\alpha^{2}$ improves in general 
the ionization energies. 
However, some inhomogeneities in their behaviour with respect to Z or N indicate that it 
is necessary to refine the value of $6s$ functions at nucleus, to accurately account 
for the finite size of the nucleus or simply to use the relativistic wave functions.

\medskip

In conclusion, the accurate studies of the structure and spectral properties of rare earth elements
require further improvement of the accounting for both the correlation and relativistic effects, but
some properties defined by valence electrons may be successfully studied by nonrelativistic approach (MCHF method) 
accounting for relativistic effects as corrections of order $\alpha^2$, 
even for heavy atoms (such as lanthanides). 

\bigskip

{\bf Acknowledgement }

The authors are grateful to C. Froese Fischer for valuable discussions and remarks.

\newpage

{\LARGE  Figures}

\bigskip
\bigskip
\bigskip
\bigskip

\begin{center}
 \includegraphics[scale=1]{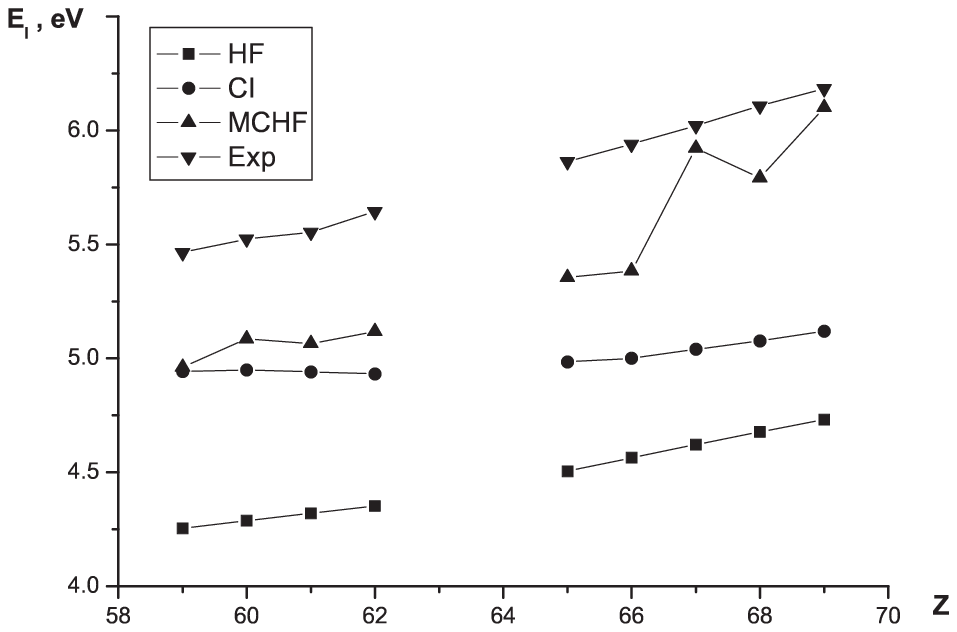}
 \end{center}
 \begin{center}
 {\bf  $Fig.$ {\it 1.} $6s$ ionization energies in various aproximations.}
\end{center}

 \begin{center}
 \includegraphics[scale=1]{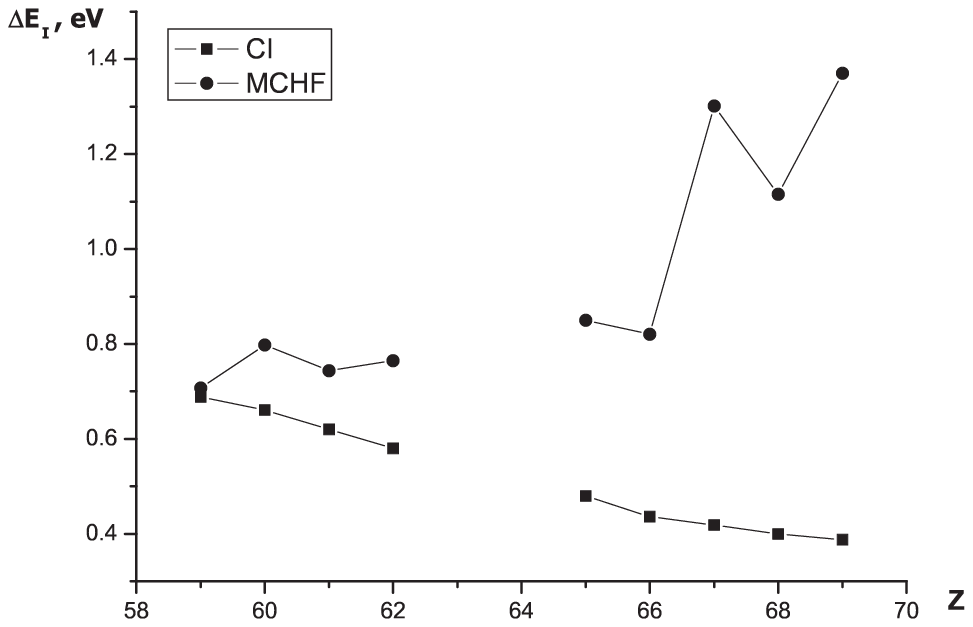}
 \end{center}
 \begin{center}
 {\bf  $Fig.$ {\it 2.} Influence of correlation effects $\Delta E_{I}$ on $E_{I}$.}
\end{center}


\newpage

{\LARGE   Tables}

\newpage

\begin{table}\centering{ \textbf{
{\large Table I. Results of single-configuration HF calculations for Er. 
Ground state energies and mean values 
of various operators in a.u. (values for Er$^{+}$ presented in brackets).}}}\\
\vspace{5 mm}
\begin{tabular}{ c c c c } \hline \hline
   $nl$    &   $<1/r>$      &   $<r>$     &   $<r^2>$    \\
\hline \hline
\\[-0.2cm]
  $1s$   &  67.45598  &  .02229 &    .00066 \\
  $2s$   &  15.76448  &  .09452 &    .01048 \\
  $2p$   &  15.76098  &  .08018 &    .00780 \\
  $3s$   &  6.01686  &  .24164 &    .06657 \\
  $3p$   &  5.94849  &  .23182 &    .06215 \\
  $3d$   &  5.84288  &  .20492 &    .04918 \\
  $4s$   &  2.55502  &  .54479 &    .33457 \\
  $4p$   &  2.45573  &  .55702 &    .35245 \\
  $4d$   &  2.24072  &  .58791 &    .40085 \\
  $4f$   &  1.72460  &  .75423 &    .73896 \\
  $5s$   &   .94798  & 1.37069 &   2.17737 \\
         &  (.93256) &(1.38534)&  (2.16005)\\
  $5p$   &   .81825  & 1.56941 &   2.80348 \\
         &  (.81981) &(1.56529)&  (2.78491)\\
  $6s$   &   .25106  & 4.63012 &  24.27349 \\
         &  (.29939) &(4.09340)& (18.75251)\\
\hline
\multicolumn{3}{l}{Energy:} & \\
\multicolumn{2}{l}{Er }     & -12498.1528   & \\
\multicolumn{2}{l}{Er$^{+}$ }& -12497.9809  & \\
\hline
\hline
\end{tabular}
\end{table}

\pagebreak

\newpage

\begin{table}\centering{ \textbf{
{\large Table II. Results of MCHF calculations. Numbers of CSFs (NCSF) and values of $E_{I}$ (in eV).}
}}\\
\vspace{5 mm}
\begin{tabular}{ c c c c c c c c c} \hline \hline
 $Basis$ & NCSF$_{Er}$ & NCSF$_{Er^{+}}$ & E$_{Er}$ (a.u.) & E$_{Er^{+}}$ (a.u.) & & $E_{I~Er}$ & $E_{I~Ho}$ \\
\hline \hline
\\[-0.2cm]
$I_{a}$ &  2838 &  2769 & -12498.58517 & -12498.38073 &   & 5.563 & - \\
[0.1cm]
$I_{b}$ & 12811 & 12054 & -12498.66977 & -12498.46502 &   & 5.572 & - \\
[0.3cm]
$II_{a}$ &   236 & 8 & -12498.17664 & -12497.96000 &   & 5.895 & 6.041 \\
[0.1cm]
$II_{c}$ &  2600 & 23 & -12498.17741 & -12497.96451 &  & 5.793 & 5.932 \\
[0.1cm]
$II_{d}$ &  5565 & 32 & -12498.17743 & -12497.96456 &  & 5.793 & 5.927\\
[0.1cm]
$II_{e}$ & 10347 & 43 & -12498.17744 & -12497.96457 &  & 5.792 &  - \\
[0.3cm]
$III_{a}$ &   70 &  4 & -12498.17657 & -12497.95988 &  & 5.896 & 19.189\\
[0.1cm]
$III_{b}$ &  272 &  7 & -12498.17729 & -12497.96428 &  & 5.796 & 5.929\\
[0.1cm]
$III_{c}$ &  733 & 11 & -12498.17733 & -12497.96446 &  & 5.792 & - \\
[0.1cm]
$III_{d}$ & 1569 & 15 & -12498.17735 & -12497.96451 &  & 5.792 & 5.923 \\
[0.1cm]
$III_{e}$ & 2938 & 20 & -12498.17735 & -12497.96452 &  & 5.792 & 5.922\\
[0.2cm]
\hline
\\[-0.2cm]
\multicolumn{3}{l}{CI \cite{SST}} & -12498.6887 & - & & 5.077 & 5.040\\
[0.1cm]
\multicolumn{6}{l}{Nonrelativistic HF \cite{SST}} & 4.677 & 4.621   \\
[0.1cm]
\multicolumn{6}{l}{Experiment \cite{MZH}} & 6.108 & 6.022  \\
[0.2cm]
\hline
\hline
\end{tabular}
\end{table}

\pagebreak

\begin{table}\centering{ \textbf{
{\large Table III. $6s$ ionization energies of lanthanides (in eV).}
}}\\
\vspace{5 mm}
\begin{tabular}{ c c c c c c } \hline \hline
 $Z$ & Atom & HF & $CI$ \cite{SST} & $E_{I}$ & Exp \cite{MZH, W} \\
\hline \hline
\\[-0.2cm]
59 & Pr & 4.254 & 4.942 & 4.961 & 5.464 \\
[0.1cm]
60 & Nd & 4.288 & 4.949 & 5.086 & 5.525 \\
[0.1cm]
61 & Pm & 4.321 & 4.941 & 5.065 & 5.554 \\
[0.1cm]
62 & Sm & 4.352 & 4.932 & 5.117 & 5.644 \\
[0.3cm]
%
65 & Tb & 4.505 & 4.985 & 5.355 & 5.864 \\
[0.1cm]
66 & Dy & 4.564 & 5.000 & 5.384 & 5.939 \\
[0.1cm]
67 & Ho & 4.621 & 5.040 & 5.757 & 6.022 \\
[0.1cm]
68 & Er & 4.677 & 5.077 & 5.792 & 6.108 \\
[0.1cm]
69 & Tm & 4.731 & 5.119 & 6.101 & 6.184 \\
[0.2cm]
\hline
$\sigma$ &  & 0.501 & 0.314 & 0.163 &  \\
[0.2cm]
\hline
\hline
\end{tabular}
\end{table}

\pagebreak

\begin{table}\centering{ \textbf{
{\large Table IV. Results of MCHF calculations of $E_{I}$  with various relativistic corrections (in eV).}
}}\\
\vspace{5 mm}
\begin{tabular}{ c c c c c c c c} \hline \hline
 $Z$ & Atom & RHF\cite{SST} &  $CI_{Est}$\cite{SST} & $E^{1}_{I}$ & $E^{2}_{I}$ & $E^{3}_{I}$ & Exp \cite{MZH, W}\\
\hline \hline
\\[-0.2cm]
59 & Pr & 4.45 & 5.24 & 5.180 & 5.180 & 5.178 & 5.464 \\
[0.1cm]
60 & Nd & 4.50 & 5.28 & 5.191 & 5.191 & 5.190 & 5.525 \\
[0.1cm]
61 & Pm & 4.54 & 5.31 & 5.242 & 5.242 & 5.240 & 5.554 \\
[0.1cm]
62 & Sm & 4.59 & 5.33 & 5.485 & 5.485 & 5.482 & 5.644 \\
[0.3cm]
65 & Tb & 4.79 & 5.45 & 5.530 & 5.528 & 5.528 & 5.864 \\
[0.1cm]
66 & Dy & 4.86 & 5.47 & 5.577 & 5.577 & 5.575 & 5.939 \\
[0.1cm]
67 & Ho & 4.93 & 5.52 & 6.686 & 6.686 & 6.680 & 6.022 \\
[0.1cm]
68 & Er & 5.00 & 5.58 & 5.878 & 5.878 & 5.877 & 6.108 \\
[0.1cm]
69 & Tm & 5.08 & 5.64 & 7.566 & 7.567 & 7.556 & 6.184 \\
[0.2cm]
\hline
 $\sigma$ &  & 0.398 & 0.152 & 0.215 & 0.215 & 0.214 &  \\
[0.2cm]
\hline
\hline
\end{tabular}
\end{table}

\end{document}